# Software and Physics Simulation at Belle II

Doris Yangsoo Kim[1]
On behalf of the Belle II Software Group

*Department of Physics, Soongsil University*
*369 Sangdo-ro Dongjak-gu, Seoul 06978, Rep. of Korea*

The Belle II experiment at the SuperKEKB collider in Tsukuba, Japan, will start physics data taking in 2018. It is planned to accumulate an $e^+ e^-$ collision data set of 50 $ab^{-1}$, about 50 times larger than that of the earlier Belle experiment. The software library for the new detector will use GEANT4 for Monte Carlo simulation and is an entirely new software and reconstruction system based on modern computing tools. Examples of physics simulation including beam background overlays will be described.

PRESENTED AT

DPF 2015
The Meeting of the American Physical Society
Division of Particles and Fields
Ann Arbor, Michigan, August 4-8, 2015

[1] Corresponding author. Tel: +82-2-820-0427; fax: +82-2-824-4383; e-mail: dorisykim@ssu.ac.kr

## 1. THE UPGRADE TO SUPERKEKB AND BELLE II

After the discovery of the Higgs particle at the LHC, the pursuit of new physics has become one of the most important tasks. The SuperKEKB/Belle II experiment is an ideal tool to look into this subject. The project can be summarized as follows: The luminosity of the $e^+ e^-$ collider, SuperKEKB, will be increased from $2.1 \times 10^{34}\,\text{cm}^{-2}\text{s}^{-1}$ to $8 \times 10^{35}\,\text{cm}^{-2}\text{s}^{-1}$, by a factor of 40 compared to that of the earlier KEKB collider. The size of the new data set will be 50 $\text{ab}^{-1}$, 50 times larger than the previous data set of 1 $\text{ab}^{-1}$[1]. The construction phase of the SuperKEKB collider is almost completed. And the construction of Belle II, the upgrade of the Belle detector, is advancing currently.

## 2. THE STRUCTURE OF THE BELLE II SOFTWARE SYSTEM

The Belle II software system, *basf2* (**B**elle **A**nalysi**S** **F**ramework 2), is a framework structure with dynamic module loading [2]. An event processing is a linear chain of modules on a path. Selection and arrangement of the modules are done by the user. *basf2* is constructed from scratch, based on the ideas from the Belle software system and other experiments such as ILC, LHCb, CDF, and Alice.

The central code management system for *basf2* is located at KEK and the revision of the library is handled by the Subversion software [3]. All common Linux operating systems are supported, including Scientific Linux and Ubuntu. The majority of code is written by C++ 11, and Python scripts are used for run steering [4].

The SCons software is used to compile and link *basf2* [5]. The automatic building of the library is done daily by a Buildbot system, with regression tests done at the same time for validation purposes [6]. An integration build of the library is created every month, and a new release of the library is conducted with emphasis on features. The Google Test, Valgrind, and Cppcheck softwares are utilized to safeguard the library from continuous activities [7-9].

Various external libraries are employed to enhance the software library. For example, ROOT is used to store common data needed for processing of events [10]. The full detector simulation is done by Geant4 9.6.2 [11], which will be upgraded to version 10 in near future. The implementation of track finding and reconstruction library is explained in Section 3.4. Calibration and alignment libraries are implemented by using the Millepede II package [12] and the "general broken lines" (GBL) fitting method [13]. The test beam data set is used to validate the package [14]. Distributed conditions databases as well as various particle identification tools are being constructed and tested. The implementation of physics analysis tools is explained in Section 3.5.

## 3. BASF2 TOOLS AND LIBRARIES

### 3.1. Geometry handling system

All the geometry parameter values are stored in the central repository in the XML format. There is a plan to move the parameters to a database later. For simulation, the actual geometry is created using C++ algorithms from the repository parameters. For the event display, the geometry is converted to ROOT TGeo using the VGM software [15].

### 3.2. Event generators

The main event generator used by *basf2* is EvtGen [16], which is interfaced with the TAUOLA, PHOTOS and Pythia 8 packages [17-19]. As the precision QED libraries, the PHOKHARA9.1, KKMC, BHWIDE, and BabayagaNLO packages are used as well [20-23]. For two-photon physics, the AAFH and KORALW packages are used [24-25]. The MADGRAPH, CRY, and Particle Gun packages are used for various purposes such as dark photon and exotic event generation, cosmic ray events, debugging, and testing [26-27].

### 3.3. Background mixing

To simulate the high luminosity environment, pre-simulated Geant4 background hits are added as SimHits (Geant4 steps) to the already created SimHits from a signal event. The background hits could be created by Touscheck, radiative Bhabha, beam-gas, or beam-wall events. Both background and signal contributions are merged and digitized at the same time.

### 3.4. Offline tracking reconstruction

The offline reconstruction of charged tracks is conducted in the following order:

1. A stand-alone finder is operated for the silicon vertex detectors (PXD and SVD) based on cellular automaton and Hopfield-Network finger [28].
2. A stand-alone finder for the central drift chamber (CDC) is operated: Legendre transformation is used to find primary tracks, while weighted cellular automaton is used for secondary tracks, decays in flight, and cosmic tracks.
3. Merge tracks found by previous two steps.
4. (Planned: Cross detector searches and extrapolations for additional hits)
5. Finally, the GENFIT Kalman Filter is applied to the track candidates delivered by the finders [29].

### 3.5. Physics analysis tools

Physics analysis is executed by loading modules to construct final state particles, a signal-side B particle, a tag-side B particle in a Python run steering script. Full event reconstruction is also possible [30]. Figure 1 shows an example Python script. The basic modules provided by *basf2* are: ParticleLoader, ParticleSelector, ParticleCombiner, and VertexFitter. Additional modules are in development to provide common analysis tools such as best candidate selection, Monte Carlo truth matching, continuum background suppression, B flavor tagging, and particle finder from the rest of the event. Also in development is the interface to Toolkit for Multivariate Data Analysis with ROOT (TMVA) [31]. Other ideas are being tested to be included as *basf2* modules as well.

### 4. SUMMARY

The Belle II experiment is a collaboration of over 600 scientists from 98 institutions in 23 countries and regions. The collaboration is actively engaging in development of a software system suitable for the next generation B factory. The basf2 framework consists of over 650k lines of C++ code (excluding comments and white space), Python scripts, and external software libraries. Many ideas have been tested and the successful ones are merged into stable library packages. Physics analysis at the end user level is possible. The library is developing at a steady pace and is expected to be operational with the coming SuperKEKB runs in 2018.

```
from basf2 import *

main = create_path( )

# optional modules for reading input data, event generation, simulation, etc.

# create final state particle lists

kaons = ('K-', '')

pions = ('pi+', '')

elecs = ('e+', '')

photons = ('gamma', '')

fillParticleLists([kaons, pions, elecs, photons], main)

# reconstruct  pi0-> gamma gamma decay

reconstructDecay('pi0 -> gamma gamma', '0.05 < M < 1.7', main)

# reconstruct D0 -> K- pi+ decay (and c.c.)

reconstructDecay(' D0 -> K- pi+', '1.6 < M < 2.0', main)

# reconstruct Btag -> D0 pi- (and c.c.)

reconstructDecay('B-:tag -> D0 pi-', '5.000 < M < 6.000', main)

# reconstruct Bsig -> pi0 e+ [nu_e] (and c.c.)

reconstructDecay('B+:sig -> pi0 e+', '0.000 < M < 6.000', main)

# reconstruct Y(4S) -> Btag Bsig

reconstructDecay('Upsilon(4S) -> B-:tag B+:sig', '0.000 < M < 11.000', main)
```

Figure 1:  An example Python steering script for physics analysis at the user level. Due to the space limit, only part of the script is shown. First, final state particles are created and listed. Then the parents of these particles are reconstructed recursively: Both the tag-side B and signal-side B particles, and Upsilon (4S) as well. In this example, the selection cut is requested on the reconstructed particle mass at each stage. Other selection methods and additional analysis utilities are available.

## Acknowledgments

This work was supported in part by Basic Science Research Program through the National Research Foundation of Korea (NRF) funded by the Ministry of Science, ICT and Future Planning (2013R1A1A3007772) and by the Supercomputing Center/Korea Institute of Science and Technology Information with supercomputing resources including technical support (KSC-2012-C1-19).